\documentclass[%
	pdftex,
	a4paper,
	oneside,
	parskip=full,
	pointlessnumbers,
	11pt,
	preprint
]{article}
\usepackage[utf8]{inputenc}
\usepackage[T1]{fontenc}
\usepackage[super,compress,numbers]{natbib}
\usepackage[left=3.4cm,right=3.4cm,top=3.4cm,bottom=3.4cm]{geometry} 
\usepackage{amsmath}
\usepackage{amsfonts}
\usepackage{amssymb}
\usepackage{placeins}
\usepackage{authblk}
\usepackage{hyperref}
\usepackage{multicol}
\usepackage[onehalfspacing]{setspace}
\usepackage{comment}
\usepackage[english]{babel}
\usepackage{graphicx}
\usepackage{tikz}
\usepackage{lineno}
\usepackage{pgfplots}
\usepackage{textgreek}
\usepackage{xcolor}

\definecolor{OliveGreen}{rgb}{0.1,0.5,0.1}

\begin{document}

\title{Mathematical modeling of variability in intracellular signaling}

\author[1,2]{Carolin Loos}
\author[1,2,3,*]{Jan Hasenauer}

\affil[1]{Helmholtz Zentrum M\"unchen-German Research Center for Environmental Health, Institute of Computational Biology, Neuherberg 85764, Germany}
\affil[2]{Chair of Mathematical Modeling of Biological Systems, Center for Mathematics, Technische Universit\"at M\"unchen, Garching 85748, Germany}
\affil[3]{Faculty of Mathematics and Natural Sciences, University of Bonn, 53115 Bonn, Germany}
\affil[*]{jan.hasenauer@uni-bonn.de}

\onehalfspacing
  \maketitle    
  
\begin{abstract}

Cellular signaling is essential in information processing and decision making. Therefore, a variety of experimental approaches have been developed to study signaling on bulk and single-cell level. Single-cell measurements of signaling molecules demonstrated a substantial cell-to-cell variability, raising questions about its causes and mechanisms and about how cell populations cope with or exploit cellular heterogeneity. To gain insights from single-cell signaling data, analysis and modeling approaches have been introduced. This review discusses these modeling approaches, with a focus on recent advances in the development and calibration of mechanistic models. Additionally, it outlines current and future challenges.
\end{abstract}

\section*{Introduction}
Cell-to-cell variability is omnipresent in biological systems and manifests itself even between genetically identical cells \cite{SpencerGau2009}. Causes include epigenetic differences, the cellular microenvironment, differences in protein expression and asymmetric cell division \cite{EbingerOez2016,AltschulerWu2010}.
This variability has been shown to influence and determine cellular decision making and has been studied in various contexts, including cancer drug resistance \cite{NiepelSpe2009}, pluripotency of stem cells \cite{FilipczykMar2015} and microbial infection \cite{MunskyKha2006}. Indeed, studies have indicated that biological systems have evolved to exploit cell-to-cell variability, for example, to enable gradual responses on the population level \cite{MitchellHof2018}.

Sources of cell-to-cell variability are often categorized as intrinsic and extrinsic noise. \citet{SwainElo2002} defined intrinsic noise as stochastic fluctuations occurring in the reaction of interest, and extrinsic noise as changes in processes influencing the reaction rates. Extrinsic noise is often assumed to be slow, resulting in stable differences between cells  \cite{RosenfeldYou2005,DunlopCox2008}, e.g., protein levels and reaction rates. 
When studying signaling pathways, the abundance of the involved biochemical species is assumed to be high enough to neglect intrinsic noise based on theory of stochastic processes \cite{vanKampen2007}.

To study variability in cellular signaling, measurements at the single-cell level are required. Experimental techniques comprise live-cell imaging using fluorescent markers \cite{Schroeder2011}, providing single-cell time-lapse data, or mass \cite{BodenmillerZun2012} and flow cytometry \cite{DaveyKel1996} for single-cell snapshot data (see \citet{GaudetMil2016} for a review on experimental techniques for single-cell signaling data). 
The data provide different types of information about the cells. Single-cell time-lapse data contain the trajectories of individual cells and are in this respect more informative than snapshot data. However, the number of measured cells and measured quantities is usually much lower. 

The complexity of signal transduction often limits the intuitive interpretation of experimental data. Therefore, various data analysis and modeling approaches have been introduced. Contributions range from statistical and information theoretical approaches, e.g., to assess the population structure \cite{SlackMar2008} and the efficiency of information processing \cite{SudermanBac2017,JetkaNie2018}, to mechanistic modeling of cellular variability and the calibration of these models (see \citet{KolitzLau2012} for a review on single-cell modeling of receptor-mediated signaling).

In this review, we outline the state-of-the-art in the mathematical modeling of cell-to-cell variability in signaling.
We discuss statistical and mechanistic models of single-cell signaling and focus on mechanistic models accounting for extrinsic noise.
We review how these mechanistic models can be calibrated to experimental data to gain insights into the biological system. We distinguish between single-cell time-lapse and single-cell snapshot data, which both pose different challenges to model calibration. In the end, we outline current and future challenges arising in modeling of cellular variability.

\section*{Mathematical modeling of single-cell signaling data}
\begin{figure}
\includegraphics[width=1\textwidth]{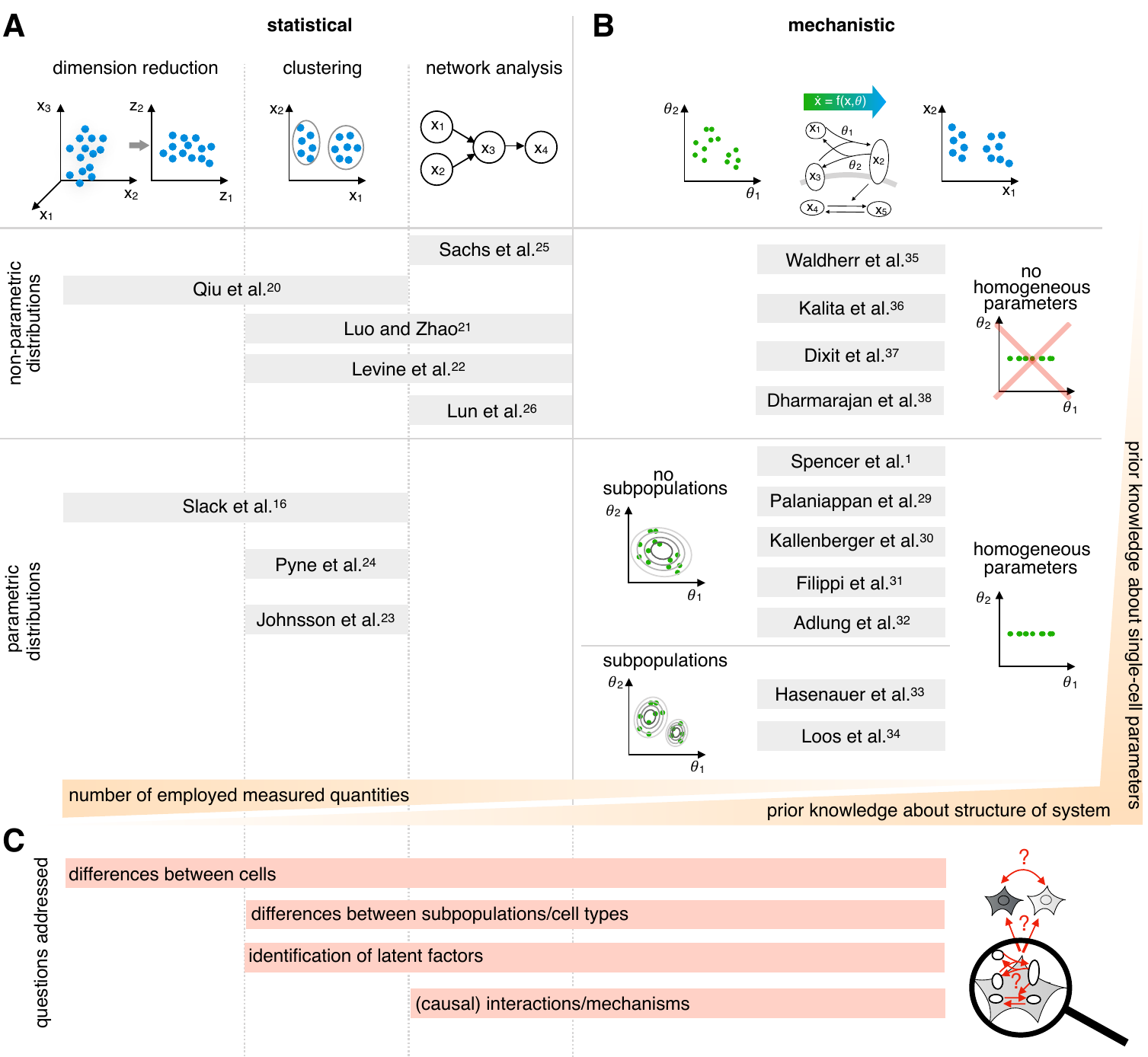}
\caption{\textbf{Overview over mathematical modeling approaches for analyzing single-cell data.}
The approaches are organized according to their class (statistical/mechanistic), the analysis tasks and their assumptions.
(\textbf{A})~Statistical modeling approaches are grouped according to their incorporation of dimension reduction, clustering or network analysis methods; and their (non-)parametric description of variability in observed cellular quantities $x_i$. (\textbf{B})~Mechanistic modeling approaches assume that variability in single-cell parameters $\theta_i$ propagates to variability in observed cellular quantities $x_i$. Methods are distinguished based on whether they employ parametric or non-parametric descriptions for the single-cell parameters, whether they account for subpopulations and whether homogeneous parameter which are shared across all cells, can be incorporated. Within a block in (A,B), publications are vertically ordered according to their publication year.
(\textbf{C})~Biological questions that have been addressed using single-cell data. The size of the bars indicates which modeling approaches, shown in (A,B), can be used to address these questions.
} 

\label{fig:overview}
\end{figure}

The choice of modeling technique highly depends on the considered biological question and the amount of available prior knowledge (Figure~\ref{fig:overview}A,B). Biological questions are often related to general differences between cells and detailed intracellular mechanisms (Figure~\ref{fig:overview}C). In the following, we discuss methods which have been used to study variability in signaling, mainly distinguishing between statistical and mechanistic models.\\

\textbf{Statistical models} are well suited to analyze large and high-dimensional data sets. A first approach is often to visualize the data in a two- or three-dimensional space, employing linear or nonlinear dimension reduction methods (Figure~\ref{fig:overview}A). These techniques, e.g., find the directions of the largest variations and thus the cellular quantities which differ most between individual cells. To study differences between subpopulations or cell types, these subpopulations/cell types need to be defined first. 
This can either be done by using established biomarkers or data-driven approaches, i.e., unsupervised clustering. Clustering is performed in the dimension-reduced \cite{SlackMar2008} or high-dimensional space of measured quantities \cite{QiuSim2011,LuoZha2011,LevineSim2015}. It is approached using a non-parametric description of the distribution \cite{QiuSim2011,LuoZha2011,LevineSim2015}, or relies on a parametric distribution for the measured quantities \cite{SlackMar2008,JohnssonWall2016,PyneHu2009}. Often, Gaussian mixture distributions \cite{SlackMar2008,JohnssonWall2016} or mixtures of alternative distributions which better cope with skewness and outliers \cite{PyneHu2009} are employed. These statistical methods have been used to, e.g., study the role of signaling variability in the response to drug treatment \cite{SlackMar2008} or determine signaling signatures in acute myeloid leukemia \cite{LevineSim2015}. If the goal is to identify interactions between cellular quantities, network based methods are often used. 
\citet{SachsPer2005} used Bayesian network reconstruction to analyze perturbation data, and reconstructed causal interactions between signaling proteins.
\citet{LunZan2017} introduced the statistical measure `binned pseudo R-squared', which is used for binned single-cell snapshot data and takes into account the deviation from the bin median over time. This facilitates the generation of hypotheses, e.g., KRAS or MEK1 being main drivers of oncogenic signaling. 
For a comprehensive review of analysis of mass cytometry data, comparing commonly used methods including SPADE \cite{QiuSim2011} and PhenoGraph \cite{LevineSim2015}, we refer to \citet{KimballOko2018} or \citet{ZielinskiThe2018}, in which the use of tools for single-cell RNA sequencing data for cytometry data is discussed. 
While these statistical methods possess rather low computation times, they cannot be used to study detailed mechanisms of the underlying biological processes.\\

\textbf{Mechanistic models} are employed to incorporate prior knowledge about the underlying biochemical reactions (Figure~\ref{fig:overview}B). This is often accomplished using ordinary differential equation (ODE) models, which describe the temporal evolution of the biochemical species. To model extrinsic noise, it is mostly assumed that parameters of the ODE model which relate to cellular properties differ between cells. In the last decade, a number of modeling approaches has been developed with different approaches to describe this property distribution. Some approaches assume parametric models for the extrinsic noise \cite{SpencerGau2009,PalaniappanGyo2013,KallenbergerBea2014,FilippiBar2016,AdlungSta2019,HasenauerHas2014,LoosMoe2018}, while others do not require such a parametric description \cite{WaldherrHas2009,KalitaSar2011,DixitLya2018,DharmarajanKal2019}. In non-parametric approaches, all parameters need to be variable, and homogeneous parameters, i.e., parameters which are shared between cells, cannot be incorporated. However, some of the approaches could be extended to allow for homogeneous parameters.

Approaches that use a parametric distribution to encode the distribution of single-cell parameters allow for homogeneous parameters.
One of the first studies which incorporated extrinsic noise in a mathematical model was performed by \citet{SpencerGau2009}, who established the importance of non-genetic variability for cellular decisions such as apoptosis. They assumed that initial protein concentrations follow a log-normal distribution and sampled from this distribution to analyze their model.
This distribution assumption was also used in other publications \cite{KallenbergerBea2014,FilippiBar2016}, while for model analysis also uniform distributions have been considered \cite{PalaniappanGyo2013}.
These methods all assume a unimodal distribution and do not explicitly account for subpopulation structures. This was addressed by \citet{HasenauerHas2014} and \citet{LoosMoe2018}. The first approach allows for differences in parameters for subpopulations \cite{HasenauerHas2014}, while the second also allows for cell-to-cell variability within subpopulations \cite{LoosMoe2018}. Accounting for subpopulation structures rendered it possible to detect that extracellular scaffolds influence nerve growth factor-induced signaling in nociceptive neurons and do not change the subgroup composition \cite{LoosMoe2018}. 

\section*{Calibration of mechanistic models}
\begin{figure}[tb]
\includegraphics[width=1\textwidth]{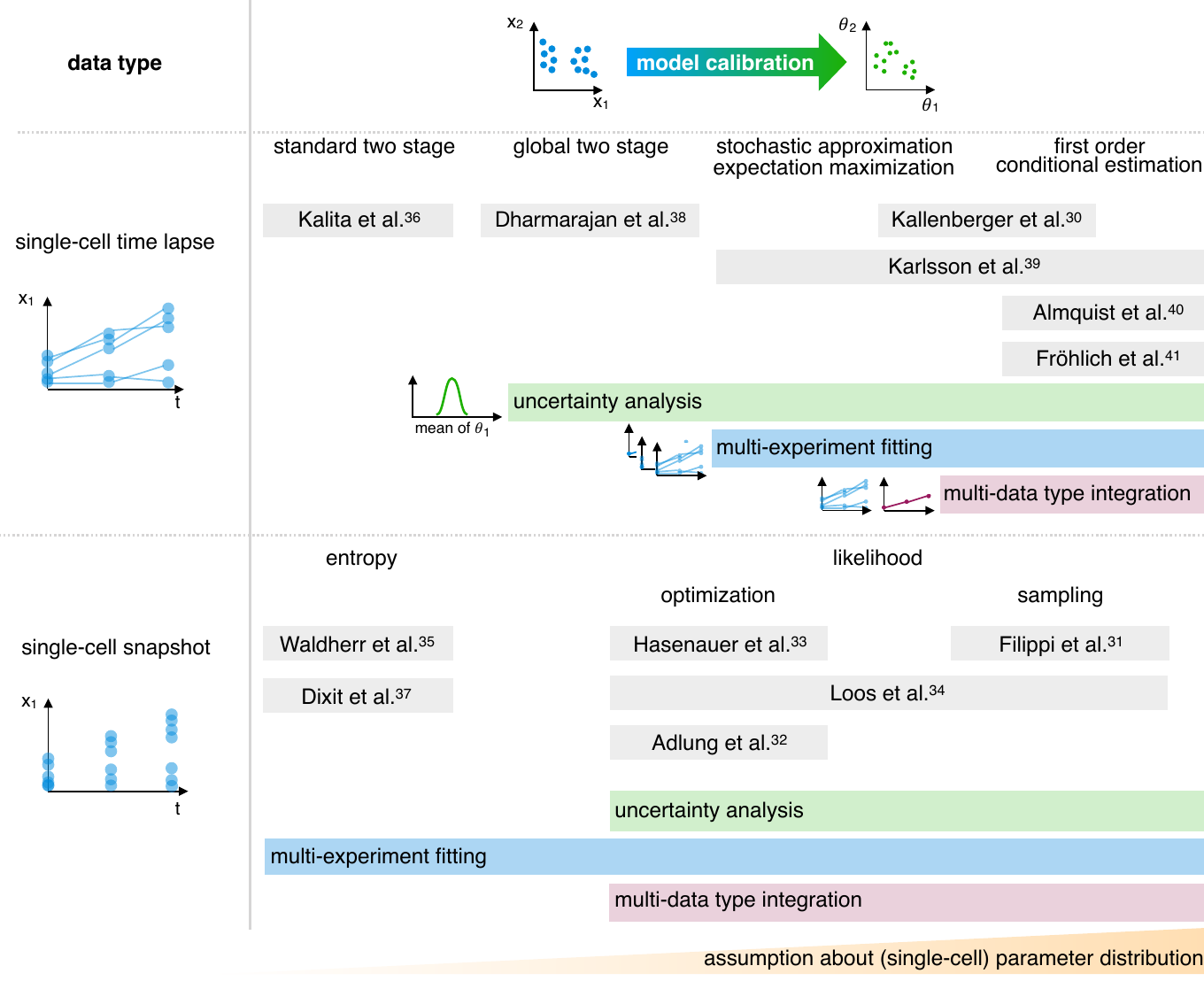}
\caption{\textbf{Single-cell data types and methods to calibrate the corresponding mechanistic models.}
Data types (left) comprise single-cell time-lapse and single-cell snapshot data, which require different model calibration techniques (right). For the different calibration approaches, corresponding publications with applications in systems biology are shown. The colored bars indicate whether the approaches allow for (i) uncertainty analysis of the estimated parameters, (ii) simultaneous fitting of multiple experiments which share parameters and (iii) the integration of different data types, e.g., population average data. 
The calibration approaches are sorted increasingly from left to right according to their assumptions made about the distribution of single-cell parameters.}\label{fig:calibration}
\end{figure}

To obtain insights into the considered biological system, the mechanistic models discussed above need to be calibrated on experimental data. Non-parametric approaches require the inference of single-cell parameters, while parametric approaches recover the distribution/population parameters, e.g., mean and covariances. For model calibration different approaches are employed depending on the use of single-cell time-lapse or single-cell snapshot data. Here, we categorize the approaches according to the following aspects: employed data type; and ability to (i)~assess the uncertainty of parameter estimates, a task which is required to distinguish between biological variability and parameter uncertainty, (ii)~simultaneously fit multiple experiments which share parameters and (iii)~integrate different data types (Figure~\ref{fig:calibration}). The integration of multiple experiments and data types facilitates the use of more information and thus can yield improved parameter identifiability. Methods which allow for more of these analyses require more assumptions about the distribution of single-cell parameters. The colored bars in Figure~\ref{fig:calibration} indicate which analyses are technically possible with the methods as they are introduced in the respective publications. However, with methodological extensions more of the analyses (i)-(iii) might be possible.
In the following, we focus on publications in the field of systems biology. Many of the methods have initially been introduced in different contexts, e.g., epidemiology or pharmacology. \\

For model fitting to \textbf{single-cell time-lapse} data, one approach is to fit the deterministic model for each cell individually and subsequently analyze the estimated single-cell parameters to obtain population parameters \cite{KarlssonJan2015}. While this approach, known as `standard two stage approach', is intuitive, uncertainty of the estimated population parameters and cellular heterogeneity cannot be distinguished and heterogeneity is often overestimated.
The global two stage approach \cite{DharmarajanKal2019} corrects for this in the second stage by exploiting a local measure for the uncertainty of single-cell parameters and an empirical Bayes estimator.
Alternative approaches integrate the two stages and consider all single-cell trajectories simultaneously. This allows the integration of different data types. \citet{KallenbergerBea2014} approximated the population likelihood by adding constraint terms to regularize for differences between single-cell parameters. In contrast, mixed-effect models marginalize out the latent single-cell parameters to obtain a population likelihood, for which no closed form is available. The first order conditional estimation (FOCE) uses a Laplacian approximation of the population likelihood \cite{AlmquistBen2015,FroehlichRei2018}, which is then maximized using gradient-based optimization. The stochastic approximation expectation maximization (SAEM) algorithm provides an iterative approximation for population likelihood and converges to the true parameter values \cite{LlamosiGon2016}.\\

For model fitting to \textbf{single-cell snapshot data}, maximum entropy approaches have been developed \cite{WaldherrHas2009,DixitLya2018}. \citet{WaldherrHas2009} discretized the observations and maximized the entropy of the parameter distribution under the constraint that measured and simulated distribution agree. \citet{DixitLya2018} improved upon this method by reformulating it to an unconstrained problem. The maximum entropy approaches do not account for measurement noise, and cellular variability and uncertainty of the parameter estimates cannot be distinguished. To distinguish these, additional statistical assumptions are made about the system. This yields a likelihood function, which either describes the whole population density \cite{HasenauerHas2014, LoosMoe2018,FilippiBar2016} or the deviance between the statistical moments of the data and those predicted by the model \cite{AdlungSta2019}. The likelihood function can then by maximized \cite{HasenauerHas2014, LoosMoe2018, AdlungSta2019}, and, e.g., subsequently be used to calculate profile likelihoods to assess the uncertainty of the estimated population parameters.
Alternatively, sampling is performed and the likelihood function is, e.g., incorporated into a Bayesian setting with a prior distribution for the population parameters \cite{FilippiBar2016,LoosMoe2018}, and uncertainty can directly be assessed from samples of the posterior distribution. Due to additional information about the parameters encoded in the prior distribution, sampling is placed further to the right in Figure~\ref{fig:calibration}, implying that it has more assumptions about the parameters. However, uniform priors can be used in a Bayesian setting or prior distributions can be incorporated in a frequentist setting. 
For efficiency, which is important for model calibration since the model needs to be evaluated many times, computationally less expensive approximations for the distributions are employed \cite{FilippiBar2016,LoosMoe2018}, e.g., sigma-point approximations \cite{Merwe2004}.

\FloatBarrier
\section*{Analysis of calibrated models}
The first step after model calibration is usually to assess whether the model is able to describe the experimental data. 
Afterwards, the parameter estimates and their uncertainty are assessed, and single-cell parameters are linked to decisions for individual cells. In particular when the model is fitted to multiple data sets simultaneously, it can unravel relations which are not apparent from the experimental data alone, e.g., if and how a factor influences a decision process \cite{HasenauerHei2012}. Sensitivity analysis can be used to predict the effects of changes in cellular properties on the behavior of individual cells. An example for this is given by \citet{AldridgeGau2011}, who identified factors for different phenotypes in receptor-mediated apoptosis using Lyapunov exponent analysis to measure the model sensitivity to initial protein concentrations. 

Many studies compare models which represent different biological hypothesis using model selection criteria, such as the Bayesian information criterion or Bayes factors, to gain a better understanding of the systems. These analyses led to insights about the contribution of intrinsic and extrinsic noise factors to variability in MAPK signaling \cite{FilippiBar2016}, dynamics of apoptosis \cite{KallenbergerBea2014}, or causal differences between neuronal subpopulations \cite{HasenauerHas2014,LoosMoe2018}.

Conceptually, the models could also be used for experimental design as done for ODE models of cellular signaling \cite{BandaraSch2009}. However, we are not aware of a publication performing this for single-cell experiments.

\section*{Challenges and outlook}
A future goal for the analysis of single-cell signaling data is to bridge statistical and mechanistic approaches, which is currently hindered by the scalability of mechanistic model calibration. When studying population average data using ODE models, methodological advances allow for the calibration of detailed models capturing a large number of biochemical species \cite{FroehlichKes2018}. The feasible dimensionality of the parameter space of mechanistic models for single-cell data is much lower, and further method development is required.

Even though it is in principle possible for many of the approaches to integrate different data types, only few publications \cite{KallenbergerBea2014,AdlungSta2019} actually used single-cell data together with population average data. Here, a key challenge is to choose the weighting of the data sets. Furthermore, the topic of uncertainty and identifiability analysis for population models inferred from different data sources needs to be addressed. While for single-cell time-lapse data established methods are applicable \cite{ChisBan2011} and for snapshot data novel approaches have been developed \cite{ZengWal2015}, it is unclear how structural identifiability from multiple data types can be addressed.

Besides methodological challenges, the use of the methods is limited by the availability of software tools. Several authors published implementations, but none of the implementations appears to be easy-to-use and flexible enough for broad use. Accordingly, the implementation of mechanistic population models and their calibration is rather time-consuming.

A challenge for all single-cell analyses is the selection and design of appropriate models for the measurement process. For flow cytometry data, it is well known that common statistical models are overly sensitive. Using measures such as the Kolmogorov-Smirnov distance, even differences between two samples under control conditions are usually significant. To avoid statistical errors and over-interpretation of data, the statistical thresholds need to be corrected \cite{Lampariello2000}. Unfortunately, in single-cell analysis this is difficult as replicates are generally not available. Advanced measurement models \cite{JohnssonWall2016} and methods to assess batch effects \cite{BuettnerMia2019} have been introduced, but it is unclear to which degree these methods address the problem.

Recent technological advances such as imaging mass cytometry \cite{GiesenWan2014} or CycIF \cite{LinFal2015} provide information about the spatial context and cell morphology. Therefore, an open question is how this information can be mechanistically exploited. The consideration of cell-cell communication would be beneficial. However, spatial models with single-cell resolution are computationally demanding and the statistical inference is challenging \cite{JagiellaRic2017}, which often requires flexible inference tools such as approximate Bayesian computation \cite{KlingerRic2018}. 

In summary, the field of model-based analysis of heterogeneous cell populations matured during the last years. There are a number of powerful modeling, simulation and calibration approaches available, and applying these to comprehensive data sets will provide novel biological insights. These approaches are urgently needed, as cellular heterogeneity and signal processing are key factors in diseases such as cancer \cite{Yaffe2019} and immunology \cite{PahlajaniAtz2011,NeuTan2017}.

\section*{Acknowledgement}
We would like to thank Lekshmi Dharmarajan, Purushottam Dixit and Benjamin M. Gyori for discussions about the methods, and Paul Stapor, Daniel Weindl, Dantong Wang, Leonard Schmiester, Elba Raimúndez-Álvarez, Yannik Schälte, Simon Merkt and Erika Dudkin for proof-reading and suggestions about the figures and manuscript. The authors acknowledge funding from the BMBF project FitMultiCell (031L0159A).
\newpage

\bibliography{Database.bib}

\end{document}